\documentclass[onecolumn]{emulateapj}
\usepackage{amsmath,amssymb,graphicx,natbib,subfigure}

\newcommand{\mh}{{M_\bullet}}
\newcommand{\msun}{{M_\odot}}

\newcommand{\beq}{\begin{equation}}
\newcommand{\eeq}{\end{equation}}
\newcommand{\sbh}{SBH}
\newcommand{\sbhs}{SBHs}

\begin{document}

\title{Gravitational Encounters and the Evolution of Galactic Nuclei. IV.
Captures mediated by gravitational-wave energy loss}
\author{David Merritt}
\affil{Department of Physics and Center for Computational Relativity
and Gravitation, Rochester Institute of Technology, Rochester, NY 14623}

\begin{abstract}
Direct numerical integrations of the two-dimensional Fokker-Planck equation are carried out
for compact objects orbiting a supermassive black hole (\sbh) at the center of a galaxy.
As in Papers I-III, the diffusion coefficients incorporate the effects of the lowest-order 
post-Newtonian corrections to the equations of motion. 
In addition, terms describing the loss of orbital energy and angular momentum due to the 
$5/2$-order post-Newtonian terms are included.
In the steady state, captures are found to occur in two regimes that are clearly differentiated in terms
of energy, or semimajor axis; these two regimes are naturally characterized as ``plunges'' (low binding energy) 
and ``EMRIs,'' or extreme-mass-ratio inspirals (high binding energy).
The capture rate, and the distribution of orbital elements of the captured objects, are presented for two 
steady-state models based on the Milky Way: one with a relatively high density of remnants and one with a lower density.
In both models, but particularly in the second, the steady-state $\overline{f}(E)$ and the distribution 
of orbital elements of the captured objects are substantially different than if the Bahcall-Wolf energy distribution were assumed.
The ability of classical relaxation to soften the blocking effects of the Schwarzschild barrier is quantified.
These results, together with those of Papers I-III, suggest that a Fokker-Planck description can adequately
represent  the dynamics of collisional loss cones in the relativistic regime.
\end{abstract}

\section{Introduction}
\label{Section:Intro}
This paper is the fourth in a series investigating the dynamical evolution of galactic nuclei containing
supermassive black holes (\sbhs).
Papers I-III \citep{Paper1,Paper2,Paper3} presented the results of time-dependent Fokker-Planck
integrations of  $f(E,L,t)$, the phase-space density of stars; 
$E$ and $L$ are respectively the orbital energy and angular momentum per unit mass.
In the Fokker-Planck description, the effects of gravitational encounters 
are represented by first- and second-order diffusion coefficients.
The diffusion coefficients in Papers I and II described changes in $E$ and $L$ due to  
 ``classical'' (random) and ``resonant'' (correlated) relaxation.
 In Paper III, the angular momentum diffusion coefficients were generalized to account for
 ``anomalous relaxation,'' the qualitatively different way in which highly eccentric orbits
 evolve in the regime, near the \sbh, where apsidal precession is due primarily to the
 weak-field effects of general relativity (GR) and is rapid enough to mediate the effects
 of resonant relaxation \citep{MAMW2011,Hamers2014}.

Still closer to the \sbh, GR implies deterministic changes in the energy and angular momentum 
of a star due to emission of gravitational waves \citep{Peters1964}.
Ordinary stars are likely to be tidally disrupted before entering this regime, but compact
objects -- white dwarves, neutron stars, and stellar-mass black holes (BHs) -- can survive
tidal disruption long enough to be captured, intact, by the \sbh. 
Capture of compact objects can occur in one of two ways: as a ``plunge,'' in which
GW emission is not important prior to the capture event; or as an ``extreme-mass-ratio
inspiral'' (EMRI), in which the object avoids capture long enough for its orbit to gradually
shrink and circularize, before spiraling into the \sbh\ \citep{SigurdssonRees1997}.

EMRIs have received much theoretical attention due to the prospect of detecting the
GW's emitted during their slow inspiral \citep{Amaro2007}.
Until now, estimates of the rate of EMRI production have been very approximate
\citep{HopmanAlexander2006,Hopman2009,Mapelli2012}.
In this paper -- for the first time -- the steady-state distribution and capture rate 
of compact objects around a \sbh\ is derived as a function of energy and angular momentum, 
including diffusion coefficients that 
account for correlated encounters and for the effects of GR up to post-Newtonian order 2.5.
In this sense, the Fokker-Planck models presented here include all of the dynamics
that were modeled in the $N$-body simulations of \citet{MAMW2011} and \citet{Hamers2014}.
However, those simulations were limited to small particle numbers ($N\lesssim 10^2$)
and short times ($\lesssim 10^7$ yr), making it difficult to extrapolate the results to
real galaxies.
The results presented here, as well as in Paper III, suggest that
the remarkable dynamics associated with the ``Schwarzschild barrier'' \citep{MAMW2011} can be captured, 
at least approximately,  via a Fokker-Planck description.
This conclusion opens the door to a much deeper understanding of the collective dynamics
of galactic nuclei in the relativistic regime.

As in Papers I-III, a single value for the stellar mass, $m_\star$, is assumed.
Since the focus of attention is compact objects, a fiducial value for $m_\star$ is
$10\msun$, typical of the stellar-mass BHs that are believed to dominate the mass-density
near the centers of dynamically-relaxed nuclei \citep{Freitag2006,HopmanAlexander2006L}.
The spin of the \sbh\ is ignored, as in Papers I-III.

Section~\ref{Section:DiffCoef} reviews briefly the method of computation and 
presents the functional forms adopted for the diffusion coefficients
in the regime where the effects of GR are important.
Section~\ref{Section:Regimes} summarizes the diffusion timescales and their dependence
on location in the phase-plane.
Sections~\ref{Section:ICs} and~\ref{Section:Results} describe the model parameters and initial conditions and
present steady-state solutions for $f(E,L)$ with parameters chosen to describe the nuclear
cluster of the Milky Way.
Section~\ref{Section:RoleSB} discusses the role that the Schwarzschild barrier plays in mediating
the capture rate of plunges in the Fokker-Planck description,
and \S \ref{Section:Summary} sums up.

\section{Diffusion Coefficients}
\label{Section:DiffCoef}

As in Papers I-III, stars are assumed to have a single mass, $m_\star$, 
and to be close enough to the black hole 
(SBH) that the gravitational potential defining their unperturbed orbits is
\beq
\Phi(r) = -\frac{G\mh}{r} \equiv -\psi(r) 
\eeq
with $\mh$ the \sbh\ mass, assumed constant in time.
Unperturbed orbits respect the two isolating integrals $E$, the energy per unit mass,
and $L$, the angular momentum per unit mass.
Following \citet{CohnKulsrud1978}, $E$ and $L$ are replaced by ${\cal E}$ and ${\cal R}$ where
\begin{eqnarray}
{\cal E} \equiv -E = -\frac{v^2}{2} +\psi(r), \ \ 
{\cal R} \equiv \frac{L^2}{L_c^2} .
\end{eqnarray}
$L_c({\cal E})$ is the angular momentum of a circular orbit of energy ${\cal E}$
so that  $0\le {\cal R}\le 1$.
${\cal E}$ and ${\cal R}$ are related to the semimajor axis $a$ and eccentricity $e$ 
of the Kepler orbit via
\beq\label{Equation:semivsE}
a = \frac{G\mh}{2{\cal E}},\ \ \ \ e^2 = 1 - {\cal R} .
\eeq
Spin of the \sbh\ is ignored.


\subsection{Anomalous Relaxation}
\label{Section:DiffCoef:AR}

In Paper III, the angular-momentum diffusion coefficients that appear in the Fokker-Planck equation
were written as
\begin{eqnarray}\label{Equation:ARDiffCoefs}
\langle \Delta{\cal R}\rangle = \langle\Delta {\cal R}\rangle_\mathrm{CK} +
g_1({\cal E},{\cal R}) \langle\Delta{\cal R}\rangle_\mathrm{RR},
\ \ \ \ 
\langle (\Delta{\cal R})^2\rangle = \langle(\Delta {\cal R})^2\rangle_\mathrm{CK} +
g_2({\cal E},{\cal R}) \langle(\Delta{\cal R})^2\rangle_\mathrm{RR}.
\end{eqnarray}
Terms with the subscript CK (``Cohn-Kulsrud'') represent classical diffusion, derived from the theory
of uncorrelated encounters \citep{Rosenbluth1957}. 
The subscript RR stands for ``resonant relaxation'' \citep{RauchTremaine1996}; these terms were 
assigned the forms
\begin{eqnarray}
\langle \Delta{\cal R}\rangle_\mathrm{RR} = 2 A({\cal E}) \left(1 - 2{\cal R}\right), \ \ \ \ 
\langle\left(\Delta {\cal R}\right)^2\rangle_\mathrm{RR} = 4 A({\cal E}) {\cal R}\left(1-{\cal R}\right) 
\label{Equation:RRDiffCoef}
\end{eqnarray}
with
\begin{eqnarray}
A(a) =  \alpha_s^2\left[\frac{M_\star}{\mh}\right]^2 \frac{1}{N} 
\frac{t_\mathrm{coh}}{P^2} , \ \ \ \ \alpha_s = 1.6.
\label{Equation:RRDiffCoef2}
\end{eqnarray}
Here $N\equiv N(r<a)$ is the number of stars instantaneously at radii less than $a$,
$M_\star = m_\star N$, $P$ is the Kepler (radial) period, and $t_\mathrm{coh}$ 
is the coherence time, defined as
\begin{eqnarray}\label{Equation:Definetcoh}
t_\mathrm{coh}^{-1} &\equiv& t_\mathrm{coh,M}^{-1} + t_\mathrm{coh,S}^{-1} \nonumber \\
t_\mathrm{coh,M}(a) &=& \frac{\mh}{Nm_\star}P\;, \ \ \ \ 
t_\mathrm{coh,S}(a) = \frac{1}{12} \frac{a}{r_g} P .
\end{eqnarray}
$t_\mathrm{coh,M}$ is the mean precession time for stars of semimajor axis $a$ due to
the distributed mass around the \sbh\ (``mass precession''), and
$t_\mathrm{coh,S}$ is the mean precession time due to the 1PN corrections to the
Newtonian equations of motion (``Schwarzschild precession'').

The functions $g_1({\cal E},{\cal R})$ and $g_2({\cal E}, {\cal R})$ account for anomalous relaxation, 
the qualitatively different way in which highly eccentric orbits evolve in response to
gravitational encounters \citep{MAMW2011,Hamers2014,BarorAlexander2014,Paper1}.
Two functional forms for $g_1$ and $g_2$ -- varying as power laws, or exponentially, with ${\cal R}$  --
were considered in Paper III. 
As noted there, numerical experiments \citep{Hamers2014}, based both on exact and restricted
$N$-body algorithms, verify the power-law forms of the anomalous diffusion coefficients and exclude
the exponential forms.
In the present paper, only the power-law forms are considered.
For $g_2$, we adopt the expression from Paper III:
\beq\label{Equation:Defineg2}
g_2({\cal E},{\cal R}) = \left\{1+\left[\frac{{\cal R}_2({\cal E})}{\cal R}\right]^n\right\}^{-2/n} , \ \ n = 8 .
\eeq
The quantity ${\cal R}_2$ is identified with the angular momentum of the Schwarzschild
barrier (SB), ${\cal R}_2\approx {\cal R}_\mathrm{SB}({\cal E})$.
Two analytic expressions have been proposed for ${\cal R}_\mathrm{SB}$:
\begin{subequations}\label{Equation:SB}
\begin{eqnarray}
\label{Equation:SB1}
{\cal R}_\mathrm{SB}^{(i)} (a) &\approx& \left(\frac{r_g}{a}\right)^2
\left[\frac{\mh}{M_\star(a)}\right]^2 N(a) ,
\\ \label{Equation:SB2}
{\cal R}_\mathrm{SB}^{(ii)} (a) &\approx& 4 \frac{r_g}{a} \frac{t_\mathrm{coh}(a)}{P(a)}\;.
\end{eqnarray}
\end{subequations}
As in Paper III, we adopt the second of these; additional reasons for doing so are presented below.
For ${\cal R}\ll {\cal R}_\mathrm{SB}$, equations (\ref{Equation:ARDiffCoefs}),
(\ref{Equation:RRDiffCoef}) and (\ref{Equation:Defineg2}) imply that the rate of
diffusion in angular momentum varies as ${\cal R}^3$:
\beq\label{Equation:Rdependence}
g_2({\cal E},{\cal R}) \langle(\Delta{\cal R})^2\rangle_\mathrm{RR} \rightarrow
4A({\cal E})\, {\cal R}_2^{-2}({\cal E})\, {\cal R}^3 .
\eeq
The low rate of diffusion at small ${\cal R}$ is a consequence of the rapid Schwarzschild
precession of eccentric orbits, which reduces the efficacy of the torques that would
otherwise induce changes in ${\cal R}$ on the resonant-relaxation timescale.

\begin{figure}[h]
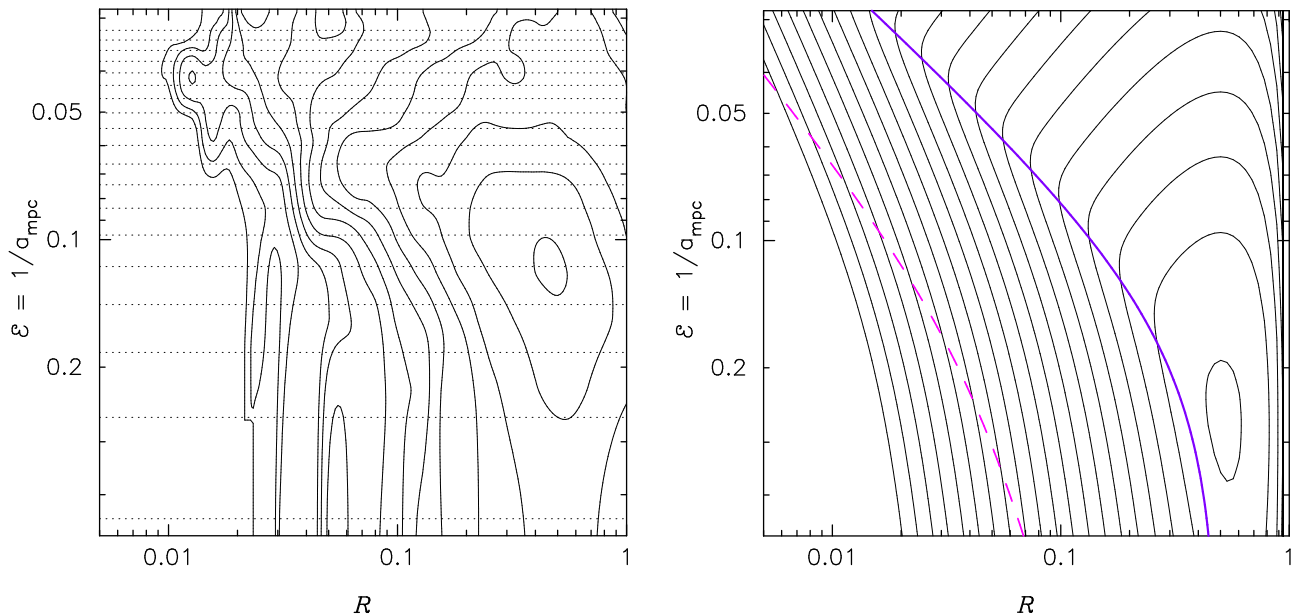

\centering
\mbox{\subfigure{\includegraphics[angle=0.,width=3.25in]{Figure1A.eps}}\quad
\subfigure{\includegraphics[angle=0.,width=3.35in]{Figure1B.eps}}\quad
}
\caption{Left panel shows contours of $\langle(\Delta{\cal R})^2\rangle$ obtained by applying
a smoothing kernel to the data of \citet{Hamers2014}.
The grid values in energy (i.e. inverse semimajor axis) are indicated by the dotted lines.
The panel on the right shows the results of fitting the model of Equation~(\ref{Equation:DiffRRfits})
to the data.
Contour spacing is the same in both panels.
The solid purple curve shows ${\cal R}_2({\cal E}, {\cal R})$.
The dashed purple curves are estimates of where classical relaxation dominates anomalous relaxation;
the analytical model would not be expected to reproduce the data to the left of this curve.
The best-fit model plotted in the right panel is shown in more detail in Figure~\ref{Figure:dfit1d}.
}
\label{Figure:dfit2d}
\end{figure}

It is interesting to make a global comparison between the expressions assumed here for the
diffusion coefficients, and the values extracted by \citet{Hamers2014} from their numerical experiments.
The left panel of Figure~\ref{Figure:dfit2d} plots $\langle(\Delta{\cal R})^2\rangle$ as found
by those authors.
Their ``field-particle'' model was designed to describe the distribution of compact remnants in the
nucleus of the Milky Way, and had these characteristics:
\begin{eqnarray}
N(<a) &=& 4.8\times 10^3 \left(\frac{a}{0.2\,\mathrm{pc}}\right),\ \ \ \ \mh=4\times 10^6\msun,\ \ \ \ m_\star=10\msun.
\end{eqnarray}
The coherence time for this model can be computed from Equation~(\ref{Equation:Definetcoh})
and is
\begin{eqnarray}
\frac{t_\mathrm{coh}}{P} \approx 4.3 \times 10^2\, {\tilde a} \left(1+0.026 {\tilde a}^2\right)^{-1} , \ \ \ \ 
P(a) \approx 1.48\, {\tilde a}^{3/2} \mathrm{yr} 
\end{eqnarray}
where ${\tilde a} \equiv a/\mathrm{mpc}$.
The energy grid adopted by Hamers et al. (indicated by the dotted lines in Figure~\ref{Figure:dfit2d}) 
was fairly coarse, and the contours
in Figure~\ref{Figure:dfit2d} were derived from the raw data using a kernel-smoothing algorithm.

The contour plot on the right of Figure~\ref{Figure:dfit2d} was obtained by adjusting the parameters 
($\alpha,\beta$) in the
expressions
\begin{subequations}\label{Equation:DiffRRfits}
\begin{eqnarray}
\langle(\Delta {\cal R})^2\rangle &=& \alpha \times 4 g_2({\cal E}, {\cal R}) A({\cal E}) {\cal R} \left( 1 - {\cal R}\right)
\\
&=& \alpha \times  4 \left\{1+\left[\frac{{\cal R}_2({\cal E})}{\cal R}\right]^n\right\}^{-2/n}  A({\cal E}) {\cal R} \left( 1 - {\cal R}\right), \\
{\cal R}_2 &=& \beta \times {\cal R}^{(ii)}_\mathrm{SB}({\cal E}) .
\end{eqnarray}
\end{subequations}
in such a way as to optimize the global fit to the data shown in the left panel.
At each energy (i.e. semimajor axis), an estimate was made of the value of ${\cal R}$ below which
classical relaxation dominates anomalous relaxation (see below), and these data were excluded
when doing the fits.

The best-fit parameters were found to be
\begin{eqnarray}
\alpha = 1.03, \ \ \beta = 1.48 .
\end{eqnarray}
These values are consistent with earlier work.
\citet{Hamers2014} computed the optimum value of $\alpha_s$ in Equation~(6) by setting
$\alpha\equiv 1$ in Equation~(\ref{Equation:DiffRRfits}); the results of our global fit confirm theirs.
And in Paper III, it was argued that a value of $\beta$ of $\sim 1.5-2$ produced the best correspondence
between the results of Fokker-Planck and $N$-body integrations.

\begin{figure}[h]
\centering
\includegraphics[angle=0.,width=6.in]{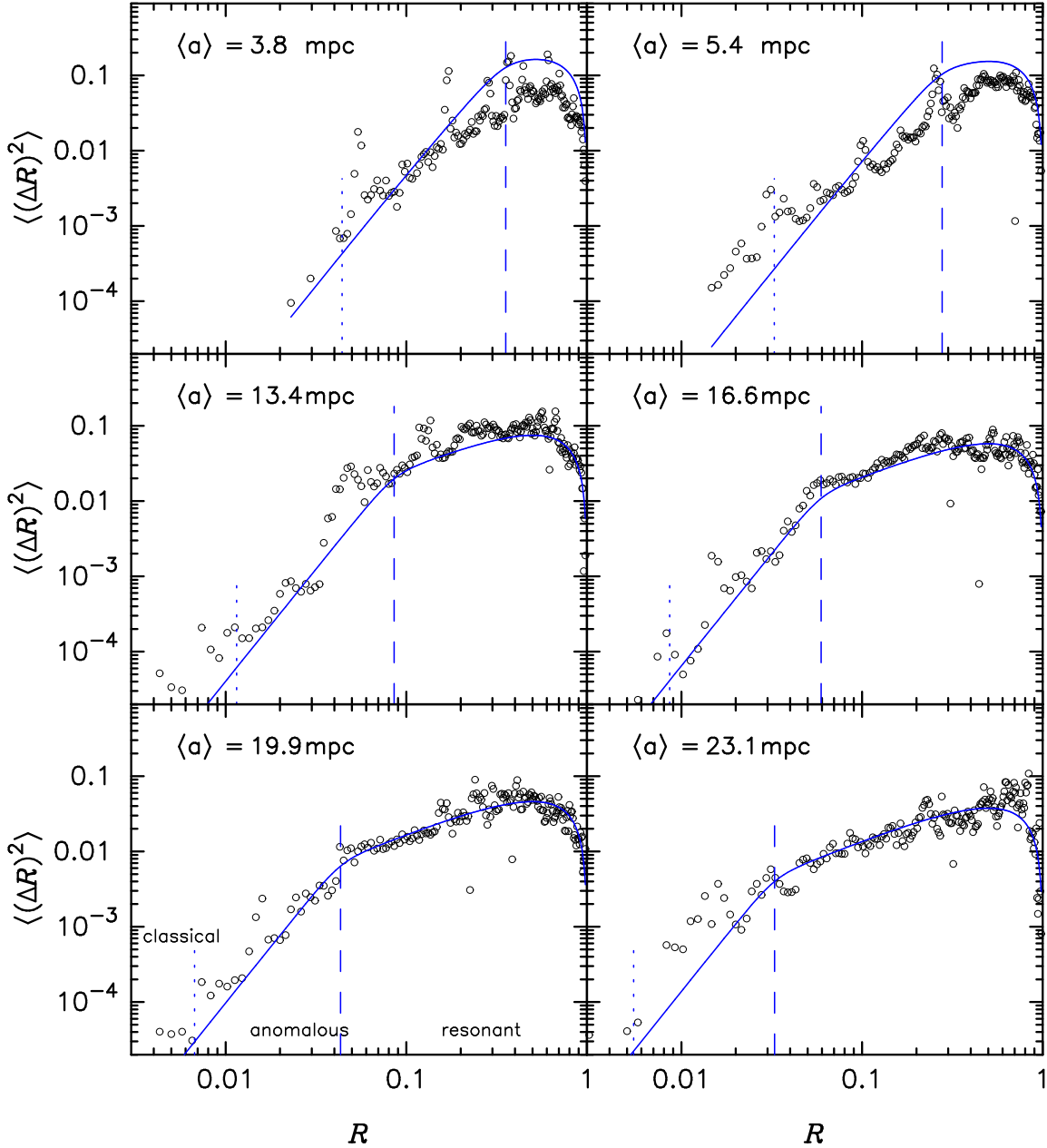}
\caption{Open circles are the data from \citet{Hamers2014} that are plotted as contours
in the left panel of Figure~\ref{Figure:dfit2d}.
Blue curves are the best-fit model shown in the right panel of that figure.
The value of ${\cal R}_2$ is shown by the dashed (blue) vertical lines.
The dotted (black) vertical lines are estimates of the value of ${\cal R}$ below which
classical relaxation dominates anomalous relaxation.
}
\label{Figure:dfit1d}
\end{figure}

Figure~\ref{Figure:dfit1d} compares model with data at several discrete energies.
Figures~\ref{Figure:dfit2d} and~\ref{Figure:dfit1d} show that the fit is not perfect: 
the model systematically over-predicts
the value of $\langle(\Delta {\cal R})^2\rangle$ at large binding energies and ${\cal R}\lesssim 1$.

As discussed in Paper III, it is natural to relate the first- and second-order
diffusion coefficients so as to satisfy a ``zero-drift'' condition, or
\begin{eqnarray}
\langle\Delta{\cal R}\rangle = \frac12\frac{\partial}{\partial {\cal R}} \langle(\Delta{\cal R})^2\rangle .
\label{Equation:DRDRRcond}
\end{eqnarray}
In Paper III, forms for $\langle\Delta {\cal R}\rangle$ were presented that contain an 
additional free parameter, ${\cal R}_1$, defined such that the zero-drift condition is satisfied
for ${\cal R}_1={\cal R}_2$.
A value $0.8\lesssim {\cal R}_1/{\cal R}_2 \lesssim 0.9$ was shown in that paper to provide
a better description of the $N$-body results of \citet{MAMW2011}.
Such values correspond to ``positive drift,'' that is, to a tendency of the particle trajectories
below the SB to move toward larger $L$.

\subsection{Gravitational wave emission}
\label{Section:DiffCoef:GW}

The integrations presented here also contain terms that describe loss of energy and angular
momentum of orbiting bodies due to emission of gravitational waves (GWs).
To lowest (2.5) post-Newtonian order, these terms are
\begin{subequations}
\begin{eqnarray}
\langle \Delta {\cal E}\rangle &=& -\frac{2{\cal E}^2}{G\mh}\langle\Delta a\rangle 
= \frac{2720}{3} \frac{m_\star}{\mh} \frac{1}{G\mh c^5}
\frac{{\cal E}^5}{{\cal R}^{7/2}} \left(1-\frac{366}{425}{\cal R} + \frac{37}{425}{\cal R}^2\right)
, \\
\langle \Delta {\cal R}\rangle &=& -2\sqrt{1-{\cal R}}\langle\Delta e\rangle =
\frac{2720}{3} \frac{m_\star}{\mh} \frac{1}{G\mh c^5}
\frac{{\cal E}^4}{{\cal R}^{5/2}} \left(1-\frac{546}{425}{\cal R} + \frac{121}{425}{\cal R}^2\right) 
\end{eqnarray}
\end{subequations}
\bigskip
\citep[][Equation~73]{Paper1}.

\section{Time scales near the loss cone}
\label{Section:Regimes}
\begin{figure}[h]
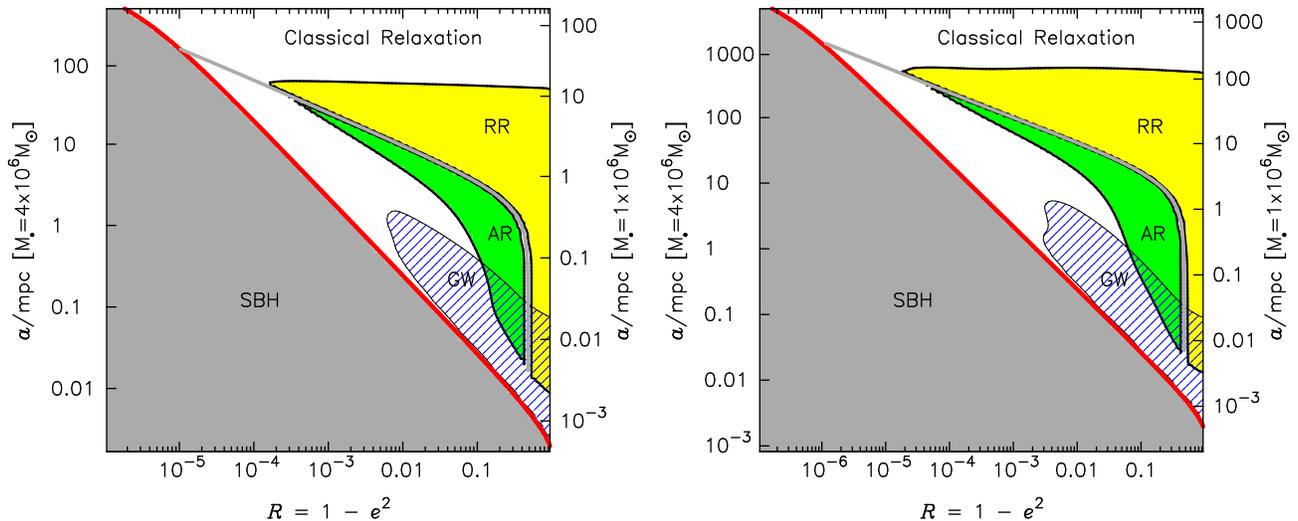

\centering
\centering
\mbox{
\subfigure{\includegraphics[angle=0.,width=3.30in]{Figure3A.eps}}\quad
\subfigure{\includegraphics[angle=0.,width=3.30in]{Figure3B.eps}}\quad
}
\caption{Comparison of time scales in two of the steady-state models from \S\ref{Section:Results}:
a ``high-density'' model (left) and a ``low-density'' model (right); these models have
$r_m/r_g\approx  \{10^7, 10^8\}$ respectively.
The vertical axis scaling assumes a SBH mass of $4\times 10^6\msun$ (left axis)
or $1\times 10^6\msun$ (right axis).
(Note the differences in the vertical range between the two panels.)
Thick red curve is the loss-cone boundary; points to the left of this curve
lie inside the \sbh.
In the white region, classical relaxation dominates the angular-momentum evolution.
In the yellow/green regions, resonant relaxation (RR) or anomalous relaxation (AR) produce changes
in ${\cal R}$ on shorter timescales than classical relaxation; the grey curve separating the
two sub-regions is the Schwarzschild barrier (SB).
Orbital evolution is dominated by gravitational wave (GW) emission in the blue hatched region;
roughly speaking,
trajectories intersecting this region are EMRIs, while those intersecting the loss-cone
boundary higher up are plunges.}
\label{Figure:regions}
\end{figure}
The various diffusion coefficients incorporated here depend in different ways on (${\cal E}, {\cal R}$) 
(or on $a$ and $e$), and so the dynamical mechanism that dominates the evolution
of orbits can be different in different parts of the phase plane.

Changes in angular momentum are dominated by random, rather than correlated, 
encounters when
\beq
\langle(\Delta {\cal R})^2\rangle_\mathrm{CK} > g_2({\cal E}, {\cal R}) \langle(\Delta {\cal R})^2\rangle_\mathrm{RR}
\eeq
(cf. Equation~\ref{Equation:ARDiffCoefs}).
The region satisfying this condition is indicated in white in Figure~\ref{Figure:regions}.
That figure was computed using two of the steady-state models described in \S\ref{Section:Results}:
a ``high-density'' model (left) and a ``low-density'' model (right).
Defining $r_m$ as the radius of a sphere containing a mass in stars of $2\mh$, the two models
have $r_m/r_g \approx \{10^7, 10^8\}$.

Interestingly, for a given density normalization -- that is, a given value of $r_m/r_g$ -- the curve
defining this region, as well as all the other regions plotted in Figure~\ref{Figure:regions}, are 
{\it independent} of $m_\star/\mh$.

The region that is the complement of the white region is divided into two sub-regions by the Schwarzschild barrier.
The (grey) curve plotted in the figure is
\beq
{\cal R}_2({\cal E}) = \beta\; {\cal R}^{(ii)}_{\mathrm{SB}} ({\cal E}),\ \ \ \ \beta = 1.5.
\eeq
As seen already in Figure~\ref{Figure:dfit1d}, anomalous relaxation can dominate the evolution
over an appreciable range in ${\cal R}$ -- more than a decade -- at many energies.
However in both of these models, classical relaxation always takes over again, at a value of ${\cal R}$
that is greater than the loss-cone value.

The blue, hatched regions in Figure~\ref{Figure:regions} 
indicate where gravitational-wave emission dominates changes in $a$ and $e$.
The boundary of this region could be defined in a number of ways.
The curve in Figure~\ref{Figure:regions} compares the time for energy loss due to GW emission:
\begin{equation}
t_\mathrm{GW}^{-1} = \left|\frac{1}{\cal E} \frac{d{\cal E}}{dt}\right|_\mathrm{GW} 
+  \left|\frac{1}{\cal R} \frac{d{\cal R}}{dt}\right|_\mathrm{GW} 
\nonumber
\end{equation}
with the timescale associated with diffusion in angular momentum from ${\cal R}$
to ${\cal R}_\mathrm{lc}$, {\bf the value defining the edge of the loss cone (Equation \ref{Equation:CaptureCond})}:
\begin{equation}
\left|\frac{\langle(\Delta {\cal R})^2\rangle}{\left({\cal R} - {\cal R}_\mathrm{lc}\right)^2}\right|^{-1} .
\end{equation}
With this definition, Figure~\ref{Figure:regions}
predicts fairly well the region in which the phase-space flow is dominated by GW emission
in the steady-state models.
But while it might be tempting to define EMRIs in terms of streamlines that cross this region,
Figure~\ref{Figure:streamlines} suggests that the streamlines are complicated enough in form that
it would be inadvisable to identify EMRIs with a particular range of initial energies or semimajor axes.

\section{Initial conditions and parameters}
\label{Section:ICs}

Steady-state solutions $f({\cal E}, {\cal R})$ were obtained via integrations of the time-dependent
Fokker-Planck equation.
As discussed in Paper I, parameters that must be specified before the start of an integration include
\beq
\frac{m_\star}{\mh}, \ \ \frac{r_\mathrm{lc}}{r_g}, \ \ \ln\Lambda, \ \ N_x,\ \ N_z :
\eeq
the stellar mass $m_\star$ (as a fraction of the mass of the \sbh, which is one in code units); 
the radius $r_\mathrm{lc}$ of the loss sphere around the \sbh, in units of $r_g\equiv G\mh/c^2$;
 the Coulomb logarithm; and the number of grid points in the $X\equiv \ln {\cal R}$ and
 $Z\equiv \ln(1+\beta{\cal E})$ directions, where $\beta$ is a constant chosen to give a judicious spacing of
 grid points at low binding energies.
In addition, the density normalization at large radius is determined by the values
$f({\cal E}_\mathrm{min}, {\cal R}_i), 1\le i \le N_x$ at $t=0$, which remain fixed during an integration.

The intent of these integrations is to model the distribution and feeding rate of stellar-mass
black holes (BHs).
In a real galaxy, the stellar BHs are only expected to dominate the total density (if they ever do)
inside a certain radius; at large distances from the galaxy center, the  remnants would
constitute at most a few percent of the mass density, depending on the stellar initial mass function 
and the galaxy's age.
The dominance of stellar remnants at small radii is an expected consequence 
of mass segregation with respect
 to the $\sim 1\msun$ stars that dominate the density at large radii in an old stellar population.
In a dynamically-evolved nucleus, and assuming classical relaxation only,
the lighter component reaches a steady-state density $\rho\sim r^{-7/4}$ at radii inside 
$r\lesssim r_\mathrm{infl}$, the \sbh\ influence radius, 
while the density of the (heavier) remnants follows a steeper, 
$\rho\sim r^{-2}$ into the radius at which they begin to dominate the density, and at smaller radii,
their density is $\rho\sim r^{-7/4}$, the single-mass Bahcall-Wolf solution
\citep[e.g.][Figure 7.7]{DEGN}.

Since the Fokker-Planck models described here contain only one mass component, 
mechanisms like mass segregation will not be present, and a decision must be made
about how to relate the single component to the population of remnants in a real galaxy.
Consider a Bahcall-Wolf single-component cusp with $\rho\propto r^{-7/4}$.
The density normalization of such a cusp can be expressed in terms of $r_m$,
the radius containing a distributed mass equal to $2\mh$.
In such a model, the mass density at a radius $r_0$ is given by
\beq
\rho(0.01\, \mathrm{pc})\approx 8.9\times 10^{17} \left(\frac{\mh}{10^6\msun}\right)^{-0.25}
\left(\frac{r_0}{0.01 \mathrm{pc}}\right)^{-1.75}
\left(\frac{r_m}{r_g}\right)^{-1.25} \msun \mathrm{pc}^{-3} .
\eeq
For the values
\beq\label{Equation:rmoverrg}
\frac{r_m}{r_g} = \{10^7, 10^8\} ,
\eeq
the implied mass density at $10^{-2}$ pc is
\begin{subequations}\label{Equation:rho0.01}
\begin{eqnarray}
\mh = 10^6\msun:\;\; \rho(0.01 \mathrm{pc}) &=& \{1.6\times 10^9, 8.9\times 10^7\} \msun \mathrm{pc}^{-3}
\label{Equation:rho0.01a}
\\
\mh = 4\times 10^6\msun:\;\; \rho(0.01 \mathrm{pc}) &=& \{1.1\times 10^9, 6.3\times 10^7\} \msun \mathrm{pc}^{-3} .
\label{Equation:rho0.01b}
\end{eqnarray}
\end{subequations}
These numbers may be compared with the density of stellar BHs predicted by steady-state, 
multi-component models for the Milky Way nucleus. 
For instance, Figure~7.8 of \citet{DEGN} suggests a density 
$\rho(0.01\;\mathrm{pc}) \approx 10^8\msun$ pc$^{-3}$ for 
the stellar ($m=10\msun$) BHs.
This value lies between the two values in Equation~(\ref{Equation:rho0.01b}).

Initial conditions for these integrations were computed as follows.
(1) An isotropic Bahcall-Wolf model, $f=f({\cal E})$,  $\rho\propto r^{-7/4}$,
was constructed having one of the two density normalizations of Equation~(\ref{Equation:rmoverrg}).
Henceforth these will be referred to as the ``high density'' ($r_m=10^7 r_g$) and
``low density'' ($r_m=10^8r_g$) models.
(2) This model was modified by setting $f$ to zero inside the loss cone, i.e., for ${\cal R}\le{\cal R}_\mathrm{lc}({\cal E})$.
The energy grid extended to ${\cal E}_\mathrm{min}\approx G\mh/r_m$ and so the
outer boundary condition consisted of a fixed density at $r\approx r_m$.
After attaining a steady state, the model densities were different at smaller radii, as discussed below,
but these changes left $r_m$ nearly unchanged.

The loss-cone boundary is defined in terms of angular momentum as
\beq\label{Equation:CaptureCond}
{\cal R}_\mathrm{lc} ({\cal E}) = 
2\frac{\cal E}{{\cal E}_\mathrm{lc}} \left(1-\frac12\frac{\cal E}{{\cal E}_\mathrm{lc}}\right),\ \ 
{\cal E} \le {\cal E}_\mathrm{lc} \equiv \frac{G\mh}{2r_\mathrm{lc}}
\eeq
where $r_\mathrm{lc}$ is the assumed radius of capture, set here to $8 r_g$, appropriate for objects
that are not tidally disrupted.
For ``plunges,'' ${\cal E}\approx 0$ and this value of $r_\mathrm{lc}$ implies $L_\mathrm{lc} \approx 4G\mh/c$.

The initial conditions just described have an empty loss cone.
In the region of most interest here, near the \sbh, the loss cone
is expected to remain nearly empty, since orbital periods are short compared with the time for
gravitational scattering to change $L$ by $L_\mathrm{lc}$ (see e.g. Figure~6 from Paper III).
Accordingly, ``empty-loss-cone'' boundary conditions --
$f({\cal E}, {\cal R})=0$, ${\cal R}\le {\cal R}_\mathrm{lc}({\cal E})$ -- were imposed,
both at initial and all later times.

As discussed above, given a value for $r_m/r_g$, the ratios between the
dynamical timescales of interest here are independent of $m_\star/\mh$.
That ratio does affect the physical unit of time however, and it was set to 
$2.5\times 10^{-6} = (10 \msun) / (4\times 10^6 \msun)$.

Following the results of Paper III, and \S\ref{Section:DiffCoef} from this paper,
the quantity ${\cal R}_2$ that appears in the expression for the second-order anomalous diffusion coefficient,
Equation~(\ref{Equation:Defineg2}), was set to $1.5 {\cal R}_\mathrm{SB}$.
The quantity ${\cal R}_1/{\cal R}_2$, which appears in the first-order diffusion coefficient,
was assigned one of the three values
${\cal R}_1/{\cal R}_2 = \{0.8,1.0,1.2\}$.
As discussed in detail in Paper III, a value ${\cal R}_1/{\cal R}_2\lesssim 1$ is required to be consistent
with the $N$-body dynamics of the Schwarzschild barrier \citep{MAMW2011}.
For ${\cal R}_\mathrm{SB}({\cal E})$, the expression (\ref{Equation:SB2}) was used.

The three choices for ${\cal R}_1/{\cal R}_2$, together with the two large-radius density
normalizations, yield a total of six models.
Scaling of a steady-state model to physical units is then determined by the value assigned to $\mh$.
Below, results are presented for two values: $\mh=\{4\times 10^6, 10^6\}\msun$.

\section{Results}
\label{Section:Results}

\begin{figure}[h]
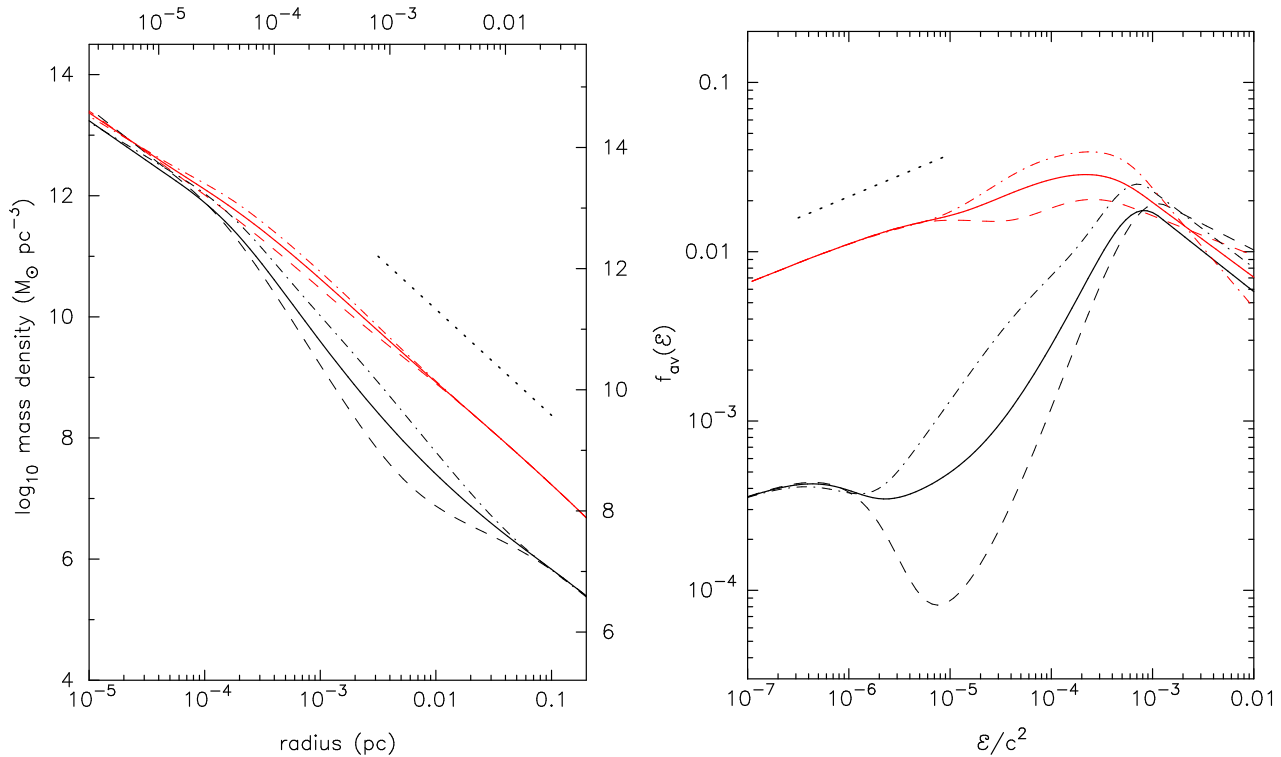

\centering
\centering
\mbox{
\subfigure{\includegraphics[angle=0.,width=3.25in]{Figure4A.eps}}\quad
\subfigure{\includegraphics[angle=0.,width=3.25in]{Figure4B.eps}}\quad
}
\caption{Mass densities (left) and angular-momentum-averaged distribution functions
(right) in the six, steady-state models.
High-density models are the red (upper) curves on both plots.
Solid curves have ${\cal R}_1/{\cal R}_2=1$; dashed curves have 
${\cal R}_1/{\cal R}_2 = 1.2$;
and dash-dotted curves have  ${\cal R}_1/{\cal R}_2 = 0.8$.
On the left-hand plot, the axis labels are based on two values for $\mh$:
$\mh = 4\times 10^6\msun$ (bottom, left axes) and
$\mh = 10^6\msun$ (top, right axes).
In the right-hand plot the dimensionless $f$ is plotted.
Dotted lines have a slope of $-7/4$ (left panel) and $1/4$ (right panel). 
}
\label{Figure:rholast}
\end{figure}

\subsection{Density and $f({\cal E})$}
\label{Section:ResultsDensity}

Steady-state density profiles are shown for the six models in the left panel of Figure~\ref{Figure:rholast}.
The right panel shows angular-momentum-averaged distribution functions, defined as
\beq
\overline{f}({\cal E}) \equiv \int_{{\cal R}_\mathrm{lc}({\cal E})}^1 f\left({\cal E}, {\cal R}\right) d{\cal R} .
\eeq
These profiles are similar to the ones shown in Figures 7 and 8 of Paper III, except at the smallest radii/largest binding energies, 
where GW energy loss dominates the evolution (Figure~\ref{Figure:regions}).
(The two values for $r_m/r_g$ adopted here correspond roughly to the two models of Paper III
having the intermediate and low densities.)
As discussed in Papers I-III, when $r_m/r_g$ is large, the steady-state density can differ substantially
from the Bahcall-Wolf solution, and that is apparent in Figure~\ref{Figure:rholast}.
While the high-density model retains the Bahcall-Wolf single-mass form, $\rho\sim r^{-7/4}$,
at most radii, the low-density model has a substantially different (steeper) slope at intermediate radii.
The corresponding $\overline{f}({\cal E})$ departs strongly from the Bahcall-Wolf single-mass form,
$f\sim {\cal E}^{1/4}$, in these models.

\subsection{Phase-space flow}
\label{Section:ResultsFlow}

\begin{figure}[h]
\centering
\centering
\mbox{
\subfigure{\includegraphics[angle=0.,width=3.75in]{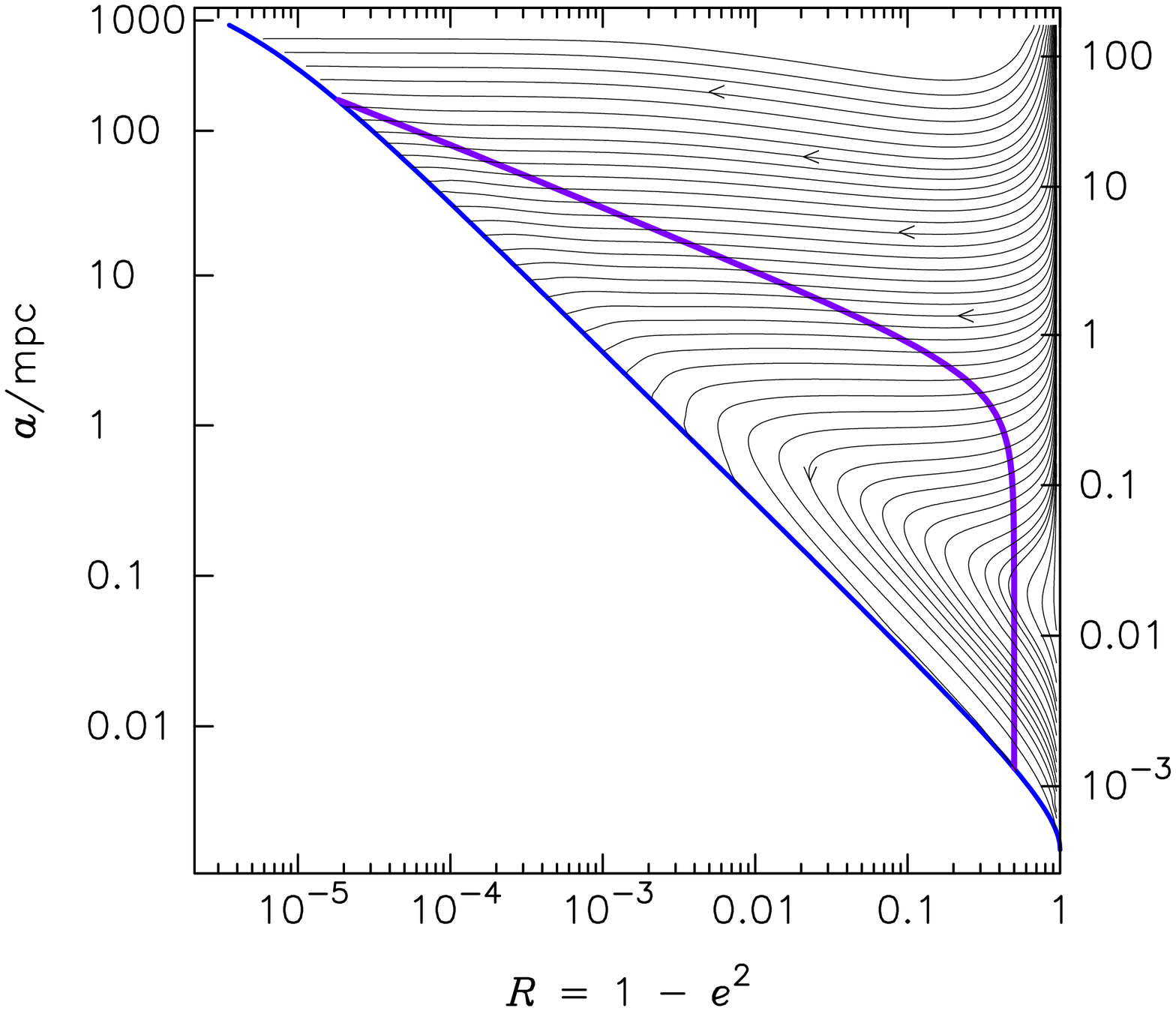}}\quad
\subfigure{\includegraphics[angle=0.,width=3.75in]{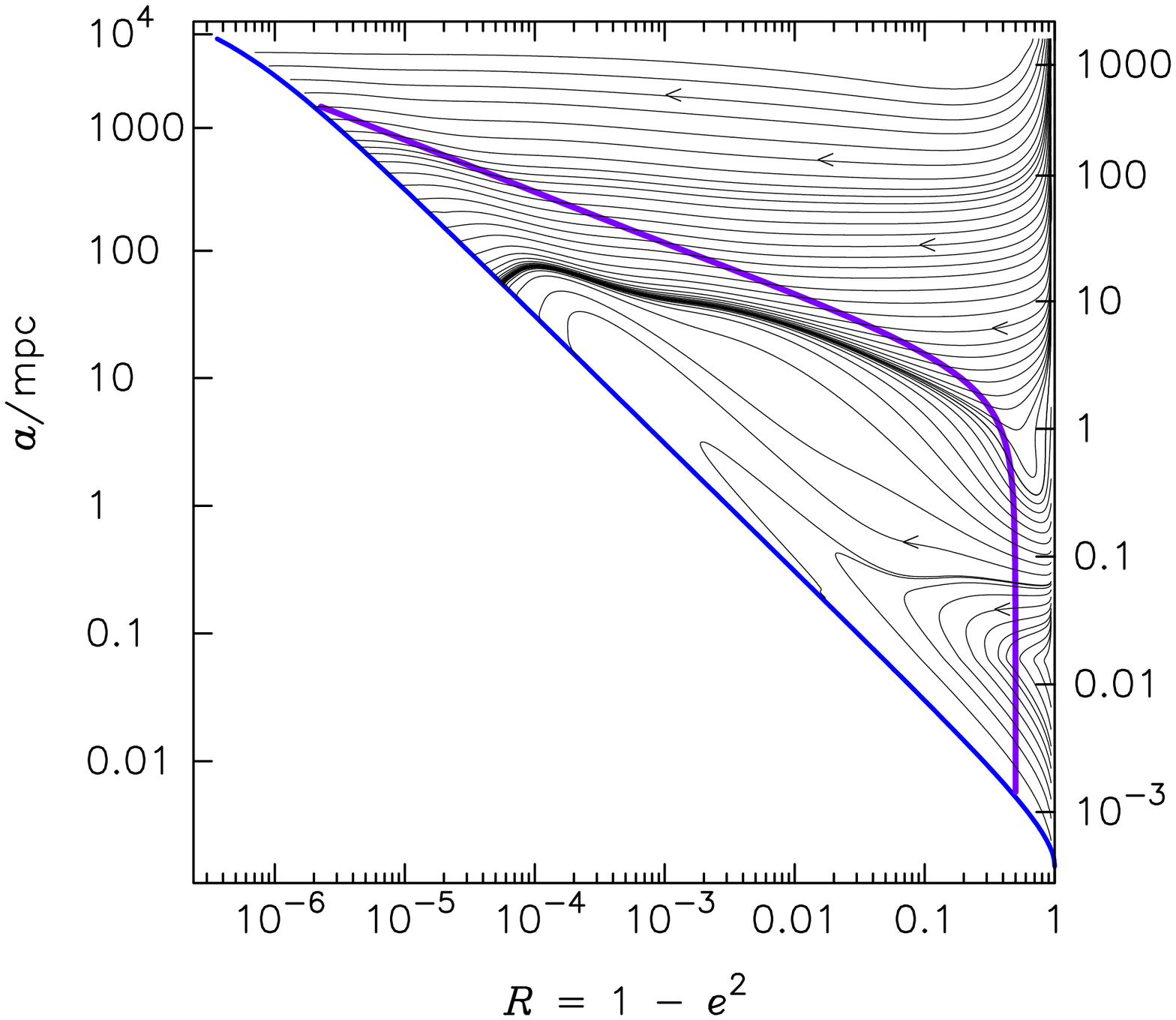}}\quad
}
\caption{Streamlines of the phase-space flux in two, steady-state models.
{\it Left:} high-density model; {\it right}: low-density model.
Both models have ${\cal R}_1/{\cal R}_2=1$.
The blue curve is the assumed loss-cone boundary and the magenta curve is ${\cal R}_2({\cal E}) =
1.5 {\cal R}_\mathrm{SB}({\cal E})$.
Vertical axes have been scaled assuming $\mh=4\times 10^6\msun$ (left axis) or
$\mh = 10^6\msun$ (right axis).
}
\label{Figure:streamlines}
\end{figure}

The mean flow of stars in (${\cal E} ,{\cal R}$) space is determined by the fluxes
\begin{eqnarray}
-\phi_{\cal E} &=& D_{\cal E\cal E}\frac{\partial f}{\partial {\cal E}} + 
D_{\cal E\cal R} \frac{\partial f}{\partial {\cal R}} +
D_{\cal E} f,\ \ 
-\phi_{\cal R} = D_{\cal R\cal E}\frac{\partial f}{\partial {\cal E}} + 
D_{\cal R\cal R} \frac{\partial f}{\partial {\cal R}} + D_{\cal R} f
\label{Equation:FPFluxConserve}
\end{eqnarray}
with flux coefficients $D_{\cal E}$, $D_{\cal E E}$ etc. given by Equations (18) of Paper I.
Streamlines of the flux for the two steady-state models having ${\cal R}_1/{\cal R}_2=1$
are shown in Figure~\ref{Figure:streamlines}.
(Streamlines in the models with different ${\cal R}_1/{\cal R}_2$ are very similar).
Comparing these plots with Figure~\ref{Figure:regions}, one sees the expected effects of the
GW loss terms near the \sbh: the streamlines become nearly parallel to the loss-cone boundary,
corresponding to a constant periapsis distance as the orbits decay.

What is not so easily predictable from Figure~\ref{Figure:regions} is the character of the
flow outside of the GW region, especially in the low-density model.
Below a certain binding energy, the streamlines in that model tend to carry stars to lower ${\cal E}$ (i.e. larger radius)
in the course of reaching the loss cone.
This is consistent with the steeper-than-$r^{-7/4}$ density  in Figure~\ref{Figure:rholast},
which tends to drive a flux toward lower ${\cal E}$ at intermediate energies.
In fact there is a ``separatrix'' streamline in this model, originating at 
$a\approx 10^3r_g$ and $e\approx 0$, that divides the flow into the two regimes:
low binding energy (${\cal E}$ initially decreases as stars are carried toward the loss cone)
or high binding energy (${\cal E}$ increases monotonically).

 
\subsection{Density in the phase plane}
\label{Section:ResultsPP}

Phase-space densities $f({\cal E}, {\cal R})$ are shown for two steady-state models in Figure~\ref{Figure:fofer}.
These models each have ${\cal R}_1/{\cal R}_2=0.8$, the preferred value.
As discussed in detail in Paper III, when ${\cal R}_1/{\cal R}_2 \lesssim 1$, the steady-state $f$ drops rapidly 
``below the SB,'' i.e. at ${\cal R}\lesssim {\cal R}_\mathrm{SB}({\cal E})$
(Figure~6 and Appendix from Paper III) and that behavior is apparent in Figure~\ref{Figure:fofer}.

\begin{figure}[h]
\centering
\centering
\mbox{
\subfigure{\includegraphics[angle=0.,width=3.35in]{Figure6A.eps}}\quad
\subfigure{\includegraphics[angle=0.,width=3.35in]{Figure6B.eps}}\quad
}
\caption{Greyscale of $f({\cal E}, {\cal R})$ in two, steady-state models.
{\it Left:} high-density model; {\it right}: low-density model.
Both models have ${\cal R}_1/{\cal R}_2=0.8$.
The blue curve is the assumed loss-cone boundary and the magenta curve is ${\cal R}_2({\cal E}) =
1.5 {\cal R}_\mathrm{SB}({\cal E})$.
Vertical axes have been scaled assuming $\mh=4\times 10^6\msun$ (left axis) or
$\mh = 10^6\msun$ (right axis).
Contrast has been set so as to highlight the drop in $f$ below the Schwarzschild barrier.
}
\label{Figure:fofer}
\end{figure}

 \subsection{Capture statistics}
\label{Section:ResultsCapture}

\begin{figure}[h]
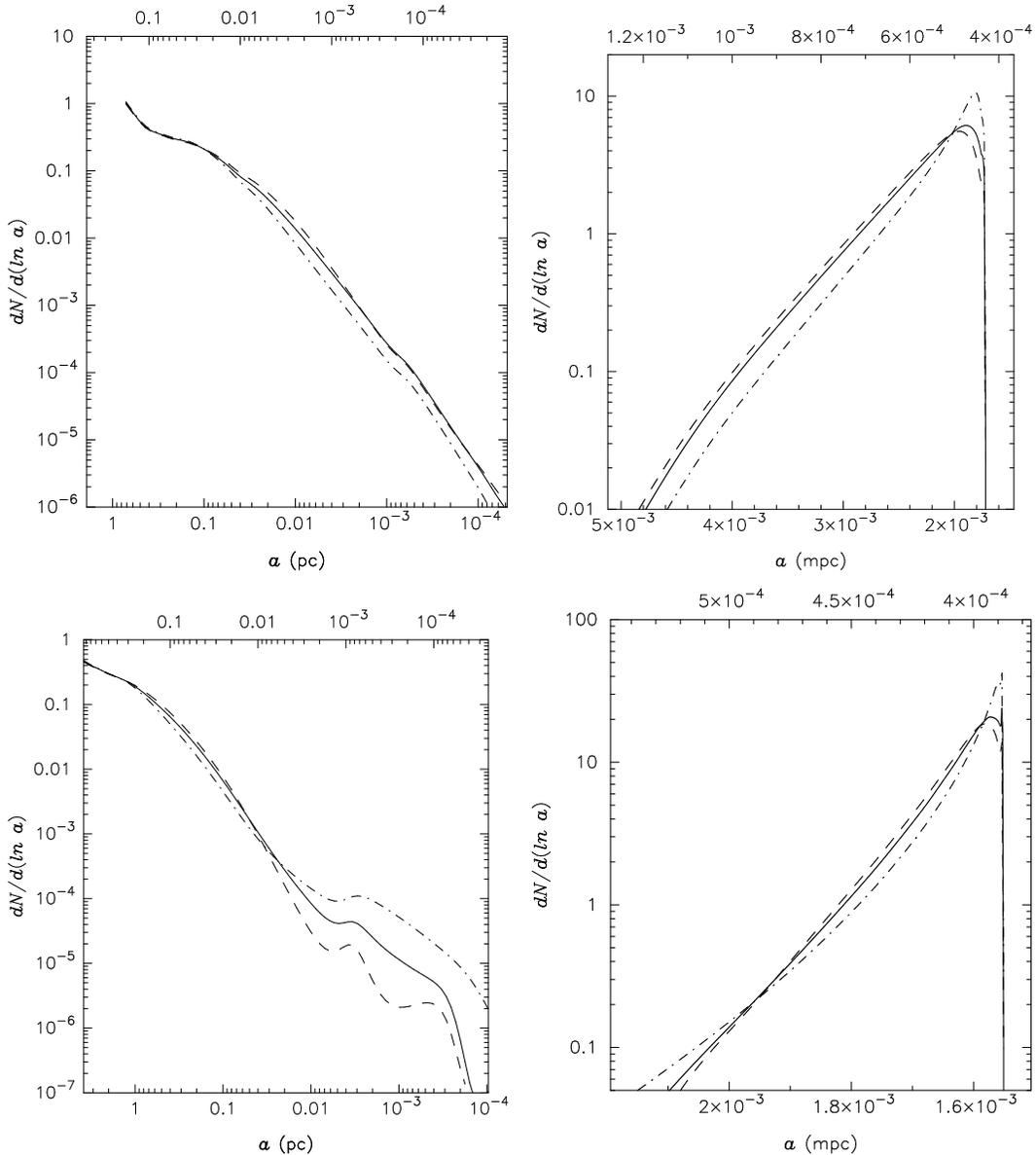

\centering
\centering
\mbox{
\subfigure{\includegraphics[angle=0.,width=2.70in]{Figure7A.eps}}\quad
\subfigure{\includegraphics[angle=0.,width=2.70in]{Figure7B.eps}}\quad
}
\mbox{
\subfigure{\includegraphics[angle=0.,width=2.70in]{Figure7C.eps}}\quad
\subfigure{\includegraphics[angle=0.,width=2.70in]{Figure7D.eps}}\quad
}
\caption{Distribution of plunge (left) and EMRI (right) events with respect to semimajor axis
at the time of capture, $a$, in the six, steady-state models.
Top frames are for the high-density models and bottom frames are the low-density models.
Line styles have the same meanings as in Figure~\ref{Figure:rholast}.
Note the very different horizontal scales on the left and right panels.}
\label{Figure:PlungeEMRI}
\end{figure}

Stars (or rather, compact objects) are lost to the \sbh\ when they cross the loss-cone boundary.
Of interest here is the total loss rate, but also its dependence on energy and angular momentum.

The rate at which stars of energy ${\cal E}$ to ${\cal E}+d{\cal E}$ cross the loss cone boundary
is easily shown to be
\begin{eqnarray}
\frac{d\dot N}{d{\cal E}} = \sqrt{2} \pi^3 {\cal E}^{-5/2}\bigg\{ 
- \phi_{\cal R}\left[{\cal E}, {\cal R}_\mathrm{lc}({\cal E})\right]
+ \frac{2}{{\cal E}_\mathrm{lc}} \left( 1 - \frac{{\cal E}}{{\cal E}_\mathrm{lc}}\right) \phi_{\cal E}\left[{\cal E}, {\cal R}_\mathrm{lc}({\cal E})\right] \bigg\} .
\end{eqnarray}
Evaluating this quantity in the steady-state models, one finds a clear separation between 
the low- and high-binding-energy regimes.
The reason, of course, is the presence of the GW terms in the Fokker-Planck equation, which
``grab'' stars at high binding energies and keep them from crossing the loss cone boundary
until their eccentricities have been sharply reduced.

Figure~\ref{Figure:PlungeEMRI} shows the energy distribution of captured stars in each of the steady-state
models.
The distributions have been plotted separately for events in the two regimes of binding energy.
It is clear from these plots that there is no {\it practical} difficulty in separating ``plunges''
from ``EMRIs'', since there are essentially no captures that take place from intermediate energies.
Note the very different ranges on the horizontal axes: capture as an EMRI always takes place from 
a (nearly circular) orbit of radius $\sim r_\mathrm{lc}\sim 8r_g$.

In the case of plunges, the distribution of orbital semimajor axes is
\beq\label{Equation:dNdaPlunge}
dN \sim a^{2/3} da,\ \ \ \ a \gtrsim 10^2 r_g
\eeq
in the high-density model.
In the low-density model, Equation~(\ref{Equation:dNdaPlunge}) is still approximately correct, although
the distribution is not so well characterized as a power-law.

Total capture rates for plunges and EMRIs in the steady-state models are given in 
Table~\ref{Table:Rates}, for the two fiducial values of $\mh$.
Plunges can occur from arbitrarily low binding energies (i.e. arbitrarily great distances).
Since these models are only valid out to $r\lesssim r_\mathrm{infl}$ --- due both to their neglect of the stellar potential,
and to the assumption of an empty loss cone, which becomes progressively more violated at low binding energies --- 
plunge rates in Table~\ref{Table:Rates} have been computed over restricted ranges in energy, corresponding
to $a\le 0.1$ pc or $a\le 0.01$ pc.

\begin{table}[ht]
\caption{Loss rates in the steady-state models}
\centering
\begin{tabular}{c c c c c c}
\hline\hline
$\mh$ & ${\cal R}_1/{\cal R}_2$ & $\rho(0.01\;\mathrm{pc})$ & ${\dot N}_\mathrm{plunge}$ (yr$^{-1}$),  &$a \le \ldots$ & ${\dot N}_\mathrm{EMRI}$ \\ 
($\msun$) & & ($\msun/\mathrm{pc}^3$) &  $0.01$ pc & $0.1$ pc & (yr$^{-1}$) \\ 
\hline 
 $4\times 10^6$& $0.8$ & $8.7\times 10^8$ & $8.4 \times 10^{-7}$ & $3.0\times 10^{-5}$ & $5.0\times 10^{-7}$ \\
                         & $1.0$ & $8.4\times 10^8$ & $1.4 \times 10^{-6}$ & $3.5\times 10^{-5}$ & $8.1\times 10^{-7}$ \\
                         & $1.2$ & $8.0\times 10^8$ & $2.0 \times 10^{-6}$ & $4.1\times 10^{-5}$ & $1.0 \times 10^{-6}$ \\ [1ex]
                         & $0.8$ & $5.9\times 10^7$ & $1.0 \times 10^{-9}$ & $9.9\times 10^{-9}$ & $2.0\times 10^{-7}$ \\
                         & $1.0$ & $2.5\times 10^7$ & $3.9 \times 10^{-10}$ & $1.4\times 10^{-8}$ & $1.8\times 10^{-7}$ \\
                         & $1.2$ & $7.5\times 10^6$ & $1.3 \times 10^{-10}$ & $1.4\times 10^{-8}$ & $2.1 \times 10^{-7}$ \\ [1ex]
 $1\times 10^6$& $0.8$ & $1.3\times 10^9$ & $1.0 \times 10^{-4}$ & $6.3\times 10^{-3}$ & $2.0\times 10^{-6}$ \\
                         & $1.0$ & $1.3\times 10^9$ & $4.8 \times 10^{-5}$ & $4.4\times 10^{-4}$ & $3.2\times 10^{-6}$ \\
                         & $1.2$ & $1.3\times 10^9$ & $6.1 \times 10^{-5}$ & $4.8\times 10^{-4}$ & $4.2 \times 10^{-6}$ \\ [1ex]
                         & $0.8$ & $4.8\times 10^7$ & $1.2 \times 10^{-8}$ & $4.8\times 10^{-7}$ & $8.0\times 10^{-7}$ \\
                         & $1.0$ & $4.1\times 10^7$ & $9.7 \times 10^{-9}$ & $6.5\times 10^{-7}$ & $7.2\times 10^{-7}$ \\
                         & $1.2$ & $4.1\times 10^7$ & $6.7 \times 10^{-9}$ & $8.4\times 10^{-7}$ & $8.3 \times 10^{-7}$ \\                         
[1ex]
\hline
\end{tabular}
\label{Table:Rates}
\end{table}

The fact that  the density profiles (Figure~\ref{Figure:rholast}) and
the flow streamlines (Figure~\ref{Figure:streamlines}) are qualitatively different in the low- and high-density models
 means that there is no simple way to scale the rates in Table~\ref{Table:Rates}
to nuclear models having other physical parameters.
But if morphological change -- i.e. evolution away from the initial conditions -- is ignored,
then loss rates (number per unit time) in these models would scale in proportion to
\beq
\left(\frac{r_m}{r_g}\right)^{-5/2} \left(\frac{c}{r_g}\right) \; \ln\Lambda \;. \nonumber
\eeq
For instance, assuming that $r_m\propto \mh\propto r_g$, the scaling of the loss rates with \sbh\ mass would be
as $\sim \mh^{-1}$.

\begin{figure}[h]
\centering
\includegraphics[angle=0.,width=3.2in]{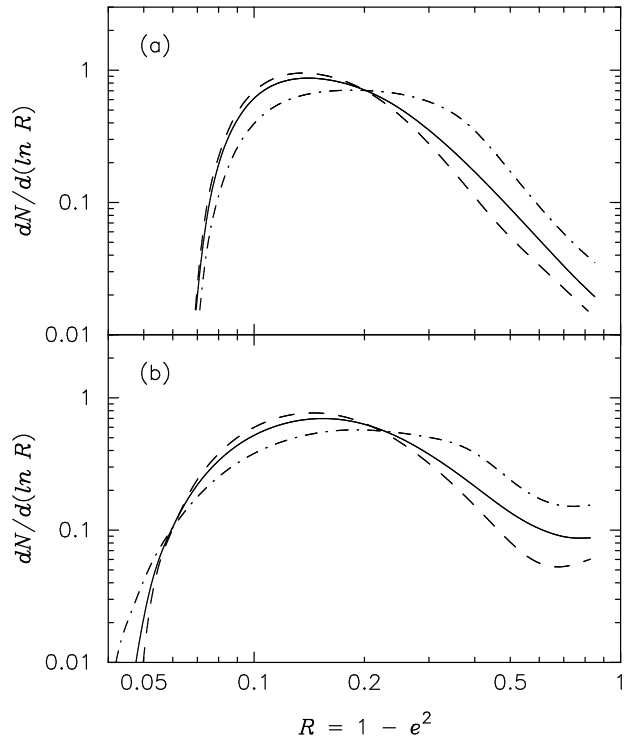}
\caption{Eccentricity distribution of the high-binding-energy (i.e. EMRI) capture events whose energy distribution
is plotted in Figure~\ref{Figure:PlungeEMRI}.
(a): High-density models; (b): low-density models.
}
\label{Figure:emrir}
\end{figure}

Figure~\ref{Figure:emrir} shows the distribution of angular momenta of the EMRIs at their moment of capture.
While capture from very eccentric orbits is rare, the peak of the distribution lies significantly away from 
circular orbits in both models, at $0.8\lesssim e \lesssim 0.9$.
\citet{HopmanAlexander2005}, using a simpler model, estimated a peak eccentricity for EMRIs of $\sim 0.7$.

\section{Classical relaxation and the role of the Schwarzschild barrier}
\label{Section:RoleSB}

The Schwarzschild barrier was discovered \citep{MAMW2011} during the course of high-accuracy $N$-body simulations,
as a locus in the ($a,e$) plane where the stellar trajectories were observed to ``bounce'' during the course of their random
walks in angular momentum.
Comparison of simulations that included, and omitted, post-Newtonian terms in the equations of motion
confirmed that the presence of the SB was associated with a substantially lower rate of captures -- or, more precisely,
of plunges, since the simulations without PN terms could not produce EMRIs.

In the Fokker-Planck description, the collisional dynamics are reduced to
whatever information can be contained in a first- and second-order diffusion coefficient.
There is no guarantee that this limited description will capture all of the remarkable dynamics exhibited
by the $N$-body trajectories.
Higher-order diffusion coefficients, or transition probabilities, might be required.
But the detailed comparison in Paper III between the $N$-body results, and solutions of the 
time-dependent Fokker-Planck equation, showed a remarkably good agreement.
Those comparisons also showed that the best correspondence was achieved if the first-order diffusion 
coefficient was allowed to differ from a ``maximum entropy'' or ``zero-drift'' form.
That possibility was allowed for here, via the parameter ${\cal R}_1/{\cal R}_2$ that appears
in the expression for $\langle\Delta{\cal R}\rangle$.

As noted in \citet{MAMW2011} and \citet{Hamers2014}, the probability of ``barrier penetration'' is also
influenced by classical relaxation, which can dominate the evolution in angular momentum if ${\cal R}$ is
sufficiently small.
The reason is that classical relaxation is unaffected by the rapid precession that quenches the effects of resonant 
relaxation.
Figures~\ref{Figure:dfit1d} and \ref{Figure:regions} from this paper suggest that classical relaxation dominates the
evolution in angular momentum when ${\cal R}\lesssim (0.1 - 1) {\cal R}_\mathrm{SB}$; the exact factor depending on the energy.

By increasing the diffusion rate at small ${\cal R}$, classical relaxation might be expected to reduce the ``hardness'' 
of the Schwarzschild barrier.
In the Fokker-Planck approximation, we can evaluate this effect as follows.
Ignoring changes in ${\cal E}$, the ${\cal R}$-directed flux is
\begin{eqnarray}\label{Equation:phiofR}
\phi_{\cal R} = - D_{\cal RR}\frac{\partial f}{\partial {\cal R}} - D_{\cal R} f 
\end{eqnarray}
with $D_{\cal R}$ and  $D_{\cal RR}$ the ``flux coefficients''  defined in Paper I.
In the case of classical relaxation, $D_{\cal R}=0$ and $D_{\cal RR} = \langle(\Delta{\cal R})^2\rangle_\mathrm{CK}/2$.
Now, it was shown in Paper II that the ${\cal R}$-dependence of the classical $\langle(\Delta{\cal R})^2\rangle$
is very close to
\beq\label{Equation:DiffRRCK}
\langle(\Delta {\cal R})^2\rangle_\mathrm{CK} = B({\cal E}) \times {\cal R}\left(1-{\cal R}\right) ,
\eeq
the same ${\cal R}$ - dependence that was adopted here for the resonant diffusion coefficient; of course
the {\it energy} dependence is very different for the two sorts of relaxation.

Let ${\cal R}_\mathrm{eq}({\cal E})$ be the value of ${\cal R}$ at which diffusion in ${\cal R}$ due
to classical relaxation, Equation~(\ref{Equation:DiffRRCK}), occurs at the same rate as anomalous diffusion.
For the latter, equation (\ref{Equation:Rdependence}), which assumes ${\cal R}\ll {\cal R}_2$,  is adequate.
We find
\beq
{\cal R}_\mathrm{eq} \approx \sqrt{\frac{B}{4A}} {\cal R}_2.
\eeq
${\cal R}_\mathrm{eq}$ is approximately equal to the value of ${\cal R}$ that is plotted as the dotted vertical
line in each of the frames of Figure~\ref{Figure:dfit1d}.
It is also roughly the value of ${\cal R}$ that defines the left boundary of the (green) regions labelled ``AR'' in 
 Figure~\ref{Figure:regions}.
Based on those figures, we expect $0.1\lesssim {\cal R}_\mathrm{eq}/{\cal R}_2 \lesssim 1$
depending on the energy.

\begin{figure}[h]
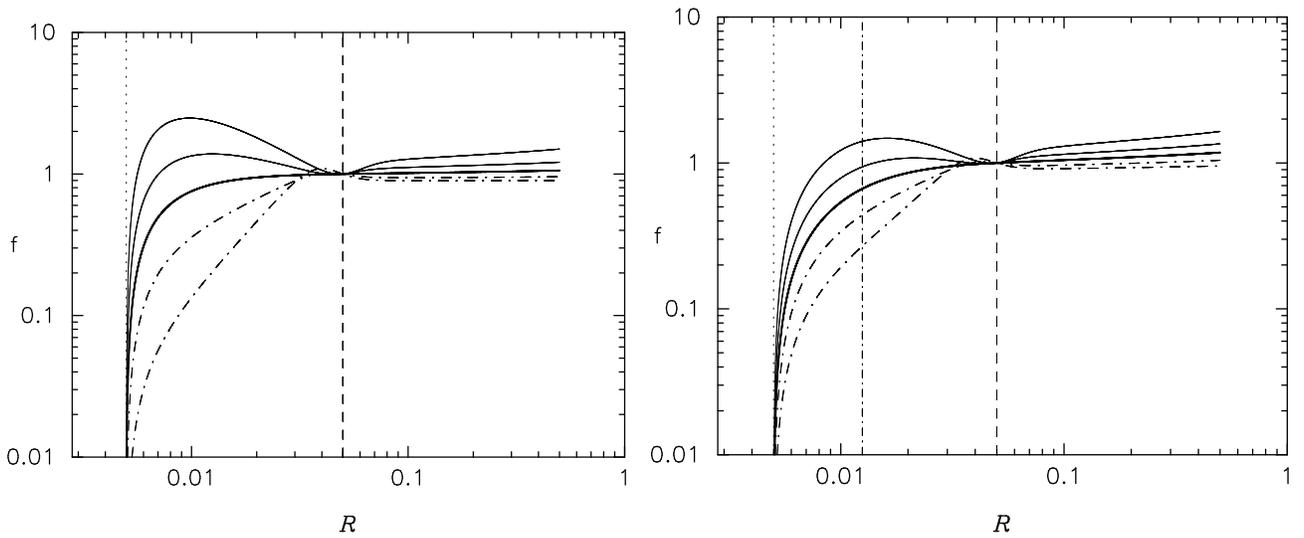

\centering
\mbox{\subfigure{\includegraphics[angle=0.,width=3.25in]{Figure9A.eps}}\quad
\subfigure{\includegraphics[angle=0.,width=3.35in]{Figure9B.eps}}\quad
}
\caption{Steady-state solutions $f({\cal R})$, computed by ignoring changes in ${\cal E}$ and
setting to zero the terms describing GW emission.
Both panels set ${\cal R}_2=0.05$ (shown by the dashed vertical line), and use the boundary
condition $f=0, {\cal R}=0.005$ (dotted vertical line).
The left panel  excludes the contribution of classical relaxation to the diffusion; this
panel is the same as Figure~10a from Paper III.
The right panel includes classical relaxation, and sets ${\cal R}_\mathrm{eq} = 0.25 {\cal R}_2$
(vertical, dash-dotted line).
For ${\cal R}< {\cal R}_\mathrm{eq}$, classical relaxation dominates the diffusion.
In both panels, the different curves have ${\cal R}_1/{\cal R}_2 = \{0.8,0.9,1,1.1,1.25\}$.
Solid lines: ${\cal R}_1/{\cal R}_2 \ge 1$; dot-dashed lines: ${\cal R}_1/{\cal R}_2 < 1$. 
}
\label{Figure:fofR}
\end{figure}

Given values for ${\cal R}_2$, ${\cal R}_1/{\cal R}_2$, and ${\cal R}_\mathrm{eq}/{\cal R}_2$,
we can solve for the steady-state $f({\cal R})$ and the flux by setting the right hand side of
Equation~(\ref{Equation:phiofR}) to a constant and requiring $f=0$ at some ${\cal R}={\cal R}_0$,
the edge of the loss cone.
(See the Appendix of Paper III for more details.)
Figure~\ref{Figure:fofR} shows the resulting $f({\cal R})$ for ${\cal R}_0=5\times 10^{-3}$,
${\cal R}_2=0.05$ and various values of ${\cal R}_1/{\cal R}_2$.
The left panel excludes the contribution from classical relaxation; this is the same figure as
in the left panel of Figure~10 from Paper III.
The right panel includes classical relaxation, with ${\cal R}_\mathrm{eq} = 0.25 {\cal R}_2$.
Accounting for classical relaxation ``softens'' the effect of barrier, causing $f$ to fall off less steeply below it.

\begin{figure}[h]
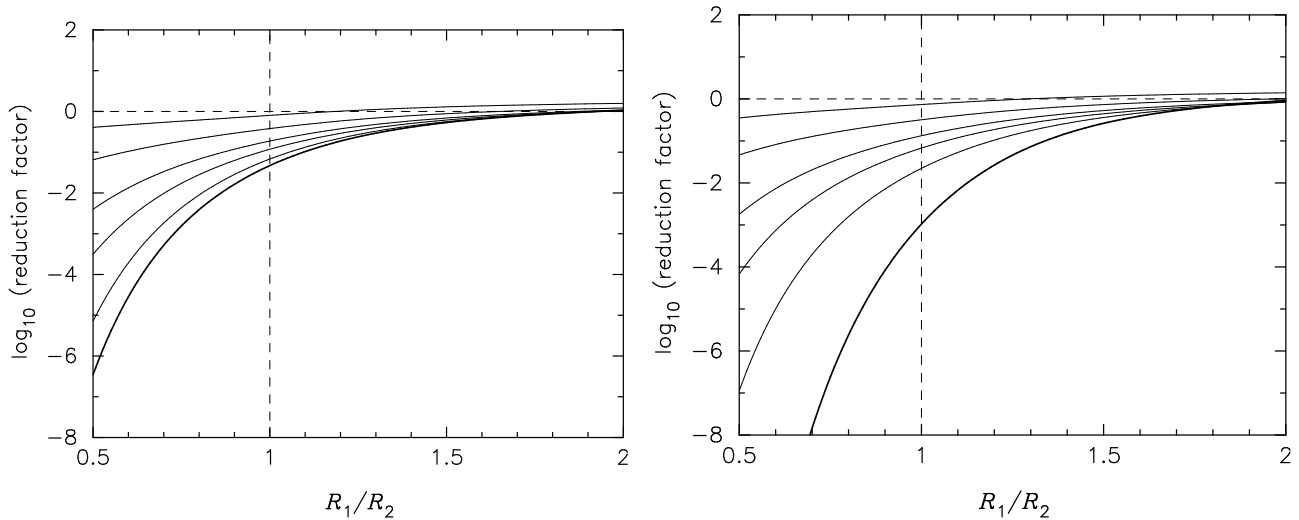

\centering
\mbox{\subfigure{\includegraphics[angle=0.,width=3.25in]{Figure10A.eps}}\quad
\subfigure{\includegraphics[angle=0.,width=3.35in]{Figure10B.eps}}\quad
}
\caption{Reduction in the steady-state flux compared to the value that would obtain if
resonant relaxation were the only mechanism causing changes in ${\cal R}$.
The left panel has ${\cal R}_2=0.05$ and boundary condition $f=0, {\cal R} = 1\times 10^{-3}$
(low binding energy).
The right panel has ${\cal R}_2 = 0.5$ and $f({\cal R}=0.01)=0$ (high binding energy).
The different curves correspond to different values of $R_\mathrm{eq}/{\cal R}_2$:
$\{0,0.1,0.2,0.3,0.5,0.8\}$, from lower to upper.}
\label{Figure:flux}
\end{figure}

Figure~\ref{Figure:flux} explores the consequences for the loss-cone flux.
Plotted there is the ``reduction factor'', defined as the ratio between the steady-state flux, and the
flux that would obtain if only resonant (not classical, or anomalous) relaxation were active.
A similar plot, excluding classical relaxation, was shown as Figure 11 in Paper III.
Figure~\ref{Figure:flux} shows that as ${\cal R}_\mathrm{eq}$ increases toward ${\cal R}_2$,
the flux reduction due to the Schwarzschild barrier becomes increasingly less severe.
{\bf  Rather than the reduction by a factor of $\sim 10^2$ (left panel) or $\sim 10^3$ (right panel) that
would obtain in the absence of classical relaxation (${\cal R}_\mathrm{eq} = {\cal R}_\mathrm{lc}$),
the actual reduction might be only a factor of $\sim 10^1$ or $\sim 10^2$, depending on energy.
}

\section{Summary}
\label{Section:Summary}

Integrations of the Fokker-Planck equation describing $f(E,L,t)$, the phase-space density
of stars around a supermassive black hole (\sbh) at the center of a galaxy, were carried out using
a numerical algorithm described in three earlier papers \citep{Paper1,Paper2,Paper3}, and including
for the first time terms describing the decay of orbits due to gravitational-wave (GW) emission.
As in Papers I-III, the functional forms chosen for the diffusion coefficients in the relativistic regime
were motivated by the results of high-accuracy numerical experiments.
Loss-cone boundary conditions were chosen to correspond to capture of compact objects by the
\sbh.
Steady-state phase-space densities, flow streamlines, and feeding rates were computed for
two models differing in degree of central concentration.
The main results follow.

\bigskip
\noindent
1. Flow of compact objects into the \sbh\ divides cleanly into two regimes depending on binding energy,
with essentially no captures taking place from intermediate-energy orbits.
Captures from orbits of low binding energy are naturally identified as ``plunges,'' 
since these orbits are relatively unaffected by GW emission prior to capture.
Captures from high-binding-energy orbits are naturally identified as ``EMRIs''
or ``extreme-mass-ratio inspirals''; for these objects, GW emission dominates the orbital evolution prior to capture.

\bigskip
\noindent
2. The semimajor axis distribution of plunge orbits is $N(a) \sim a^{2/3} da$.
All of these orbits are highly eccentric.
Capture of EMRIs takes place from orbits near in energy to that of the most-bound orbit around the \sbh,
but their eccentricity distribution is fairly broad, $0 \leq e \lesssim 0.97$, with a peak near
$e=0.9$.

\bigskip
\noindent
3. Steady-state capture rates were computed for plunges and EMRIs in two nuclear models with low and high
degrees of central concentration.
Scaled to the mass of the Milky Way \sbh, EMRI rates in the high(low)-density models are $\sim 5(2) \times 10^{-7}$ yr$^{-1}$.
The rate of plunges, from orbits with semimajor axes less than $0.1$ pc, is 
$\sim 3\times 10^{-5}$($\sim 1\times 10^{-8}$) yr$^{-1}$ in the two models.
Approximate relations were presented for scaling the rates to other nuclei.

\bigskip
\noindent
4. Results obtained here, and in Paper III, demonstrate that the remarkable dynamics associated with
the Schwarzschild barrier can be captured in a Fokker-Planck description, a result that opens the door
to a deeper understanding of the consequences of general relativity for the long-term collisional 
evolution of galactic nuclei.


\acknowledgements
I thank A. Hamers for providing data from his {\tt TPI} code that was used in constraining
the functional forms of the anomalous diffusion coefficients in \S \ref{Section:DiffCoef}.
This work was supported by the National Science Foundation under grant no. AST 1211602 
and by the National Aeronautics and Space Administration under grant no. NNX13AG92G.

\end{document}